\documentclass[12pt]{article}

\renewcommand{\title}[1]{\begin{center}\bf\Large #1\end{center}}

\renewcommand{\author}[1]{\begin{center}\large #1\end{center}}

\usepackage{latexsym,amsfonts}
\newcommand{\rr}{\mathbb{R}}
\newcommand{\sll}{SL(2,$\rr$)}
\newcommand{\slu}{SL(2,$\rr$)/U(1)}
\newcommand{\slua}{sl(2,$\rr$)}
\newcommand{\tr}{{\rm tr}}

\newcommand{\intl}{\int\limits}

\newcommand{\pk}[2]{\left\{#1,#2\right\}}
\newcommand{\del}{\partial}
\newcommand{\delz}{\partial_z}
\newcommand{\delzb}{\partial_\zb}
\newcommand{\ub}{{\bar{u}}}
\newcommand{\zb}{{\bar{z}}}

\newcommand{\phib}{{\bar{\phi}}}

\begin{document}

\title{Integration of the \slu{} Gauged WZNW Theory by
   Reduction and Quantum Parafermions}
 \author{C. Ford${}^a$,
 G. Jorjadze${}^b$
 and G. Weigt${}^a$ \\
{\small${}^a$DESY Zeuthen,
    Platanenallee 6,}\\{\small D-15738 Zeuthen, Germany}\\
{\small${}^b$Razmadze Mathematical Institute,}\\
  {\small M.Aleksidze 1, 380093, Tbilisi, Georgia}}

\begin{abstract}
  Using a gauge invariant reduction we directly integrate the \slu{}
  WZNW theory. We prove that the conserved parafermions of this
  theory are coset currents.  Quantum mechanically, the parafermion
  algebra, the energy-momentum tensor, and `auxiliary' parafermions
  are deformed in a self-consistent manner.
\end{abstract}

\baselineskip=20pt

\section{Introduction}

Gauged Wess-Zumino-Novikov-Witten (WZNW) models are an important
subclass of two-dimensional integrable conformal field theories. But
there is an essential structural difference between nilpotent
\cite{Balog} and non-nilpotent \cite{BCR}
 gauging. Although the equations of motion for both cases
 have a linear Lax pair representation \cite{CMP},
 only the Lax pairs for nilpotently gauged (Toda) theories
have been integrated directly \cite{LS}.

In this paper we will focus on non-nilpotently gauged WZNW theory. In
contrast to the Toda case a systematic integration method is lacking.
Instead of integrating a Lax pair, we extend the methods of Lagrangian
and Hamiltonian reduction in a gauge invariant manner and prove that a
non-nilpotently gauged WZNW theory can be integrated directly.  We
rederive here the general solution of the \slu{} theory, which was found
in refs \cite{MuWe,NuPh99} in a non-systematic manner guided by the
solution of the $B_2$ non-abelian Toda theory \cite{GS,Bilal}.  While
we will restrict ourselves to the simplest \slu{} case, we presume
that this approach can be generalised to any gauged WZNW theory.

Gauge invariant reduction also proves that the
parafermionic conserved quantities are the coset currents,
and the reduced
energy-momentum tensor retains the simple Sugawara form in terms of
these coset currents.

The \slu{} theory is not only of mathematical interest. This model
attracted much attention when it was realised that such theories have
a black hole interpretation \cite{Witten}. In the early 1990's studies
of the quantum \slu{} WZNW theory relied heavily on rather formal path
integral manipulations \cite{Witten,BS,Tseytlin} or related operator
identities \cite{DVV}. Since the attempted path integration over the
$U(1)$ gauge field provided an incomplete effective action
\cite{CMP,Buscher}, we define the \slu{} theory by the classical
Lagrangian given in ref. \cite{BCR}. Our goal is to perform for this
theory an exact canonical quantisation, much in the same way as it has been done
for Liouville theory \cite{BCT,OW}. Here one recasts the general
solution as a canonical transformation exchanging interacting and free
fields and lifts the classical conformal transformation to an
operator transformation.

We take the parafermion algebra as a starting point for quantisation.
In doing this we are forced to deform the classical free field
representation of the parafermions. Similar deformations have been
found before in OPE based Feigin-Fuks constructions of WZNW Kac-Moody
currents \cite{FZ,Ger,JNS,Nem,Bakas}.  In addition we derive the
 quantum analogue of the parafermion algebra. The Sugawara
form of the energy-momentum tensor motivates us to build the quantum
energy-momentum tensor using only the parafermions, and we find an
improvement term which in the $\sigma$-model picture could
correspond to a non-perturbative dilaton. Finally, we point out that
the general solution of the model has a parafermion interpretation.
More precisely, the general solution contains fields of conformal
weight zero related to an alternative set of `auxiliary' parafermions
which
also undergo quantum deformation.

\section{ A Lagrangian Reformulation of the
 \sll \\  WZNW Theory}

WZNW models are defined by the action \cite{WZ,N,W}
\begin{equation}\label{WZNW-Wirkung}
S_{\rm{WZNW}}[g]=\frac{k}{8\pi}\intl_M h^{\mu\nu}
\tr\left(g^{-1}\del_\mu gg^{-1}\del_\nu g\right)
\sqrt{-h}\,\, d\sigma\, d\tau +kI_{\rm{WZ}}[g],
\end{equation}
which includes the topological Wess-Zumino term
\begin{equation}\label{WZ-term}
  I_{\rm{WZ}}=\frac{1}{12\pi}\intl_{B}\tr\left( g^{-1}\mbox{d} g\wedge
g^{-1}\mbox{d}
  g\wedge g^{-1}\mbox{d} g\right).
\end{equation}
Here $h_{\mu\nu}=diag(+,-)$ is the Minkowskian metric of the world
surface $M$, $h$ its determinant, $B$ a volume with the boundary
${\partial B=M}$, $k$ the coupling parameter, and $g(\tau,\sigma)$ is
a field  which takes values in a
semi-simple
Lie group $G$.
 We shall restrict ourselves in this paper to
the case $G=$ \sll.

To simplify the construction of the \slu{} theory it will
prove useful  to rewrite the
topological WZNW
term (\ref{WZ-term}) as an integral of a local Lagrangian with global $U(1)$ symmetry.
We are going to verify
that the differentiation of the 2-form (for a more general treatment
see \cite{FJW})
\begin{equation}\label{2-form+}
 F =\frac{2}{1 + \langle a~g~a~g^{-1}\rangle }\,
 L_{a}\wedge R_{a}
\end{equation}
provides the integrand of the topological WZ term (\ref{WZ-term})
\begin{equation}\label{3-form}
dF = \frac{2}{3}\langle g^{-1}dg \wedge g^{-1}dg \wedge g^{-1}dg \rangle,
\end{equation}
and then  Stokes' theorem  reduces the Wess-Zumino term
 to a two dimensional integral of $F$ over $M=\partial B$.
Here $a$ is a fixed normalised time-like element of the \slua \, algebra
with $\langle a~a\rangle=1$, where $\langle \cdot \rangle=-\frac{1}{2}\mbox{tr}
(\cdot )$ denotes a  normalised trace, and the left and right 1-forms of
(\ref{2-form+}) are given by
\begin{equation}\label{1-forms}
 L_a =\langle a~dg~g^{-1}\rangle ,~~~~~~
 R_a =\langle a~g^{-1}dg\rangle .
\end{equation}
Let us introduce the
basis of the \slua \, algebra
\begin{equation}\label{T}
  t_0=\left( \begin{array}{cr}
  0&-1\\1&0 \end{array}\right),~~~~
   t_1=\left( \begin{array}{cr}
  0&~1\\1&~0 \end{array}\right),~~~~
 t_2=\left( \begin{array}{cr}
  1&0\\0&-1 \end{array}\right).
\end{equation}
It satisfies the relations
\begin{equation}\label{TxT}
t_m~t_n=-\eta_{mn}~I+\epsilon_{mn}~^l~t_l,
\end{equation}
where $I$ is the unit matrix,
$\eta_{mn}=\mbox{diag}(+,-,-)$ the $3d$
Minkowskian metric, and
$\epsilon_{012}=1$. The normalised traces of the matrices $t_n$ are
then given by
\begin{equation}\label{trace}
\langle t_m~t_n \rangle =\eta_{mn},~~~~~~~
\langle t_l~t_m~t_n \rangle =\epsilon_{lmn}.
\end{equation}
This defines an
isometry between the \slua \, algebra and  $3d$ Minkowski space.

The left and right 1-forms
\begin{equation}
 L_n =\langle t_n~ dg~g^{-1}\rangle ,~~~~~~
 R_n =\langle t_n~ g^{-1}~dg\rangle
\label{1-form}
\end{equation}
 are related by $ L_m =\Lambda_{m}^{~~n}(g) R_n $, where
\begin{equation}
\Lambda_{m}^{~~n} (g)=\langle t_m~ g~t^n~g^{-1} \rangle
\label{Lambda_mn}
\end{equation}
is a Lorentz transformation matrix. Since
from (\ref{1-form}) we have
$~g^{-1}~dg = t^n R_n~$ and $~dg~g^{-1} = t^n L_n~$,
using (\ref{trace}) the right hand side of
(\ref{3-form}) can be written in terms of right (or left) 1-forms
\begin{equation}
\frac{2}{3}\langle g^{-1}dg \wedge g^{-1}dg \wedge g^{-1}dg \rangle
=4~ R_0\wedge R_1\wedge R_2.
\end{equation}
Moreover, the differentials of (\ref{1-form}) and (\ref{Lambda_mn})
\begin{eqnarray}
d L_n& =&\epsilon_n^{~~lm}~L_l\wedge L_m,~~~~~~~
d\Lambda_{mn}=2\epsilon_n\ ^{kl}~\Lambda_{mk}~R_l
\nonumber \\
d R_n &=&-\epsilon_n^{~~lm}~R_l\wedge R_m,
\end{eqnarray}
give for the left hand side of (\ref{3-form})  the same result
$4~ R_0\wedge R_1\wedge R_2$. This proves our statement
(\ref{3-form}). With Stokes' theorem, we finally obtain the alternative
Lagrangian formulation of the WZNW theory
\begin{equation}\label{L}
S=\int_M dz d\bar z ~ {\mathcal L}, ~~~~~~~~
 {\mathcal L}={\mathcal L}_0 + {\mathcal  L}_{WZ},
\end{equation}
where the kinetic term ${\mathcal L}_0$ remains unchanged
\begin{equation}\label{L_0}
{\mathcal L}_0= -\frac{1}{\gamma^2}~
\langle g^{-1}\partial_z g~ g^{-1}\partial_{\bar z}g \rangle ,
\end{equation}
but the WZ term becomes
\begin{equation}\label{L_{WZ}}
{\mathcal L}_{WZ} =-\frac{1}{\gamma^2}\,
\frac{\langle a~\partial_zg~ g^{-1} \rangle
\langle a~ g^{-1}\partial_{\bar z}g\rangle -
\langle a~\partial_{\bar z}g~ g^{-1} \rangle
\langle a~ g^{-1}\partial_zg~\rangle}
{1+\langle a~g~a~g^{-1}\rangle} .
\end{equation}
Here we used light-cone coordinates $z=\sigma +\tau,~ \bar z=\tau
-\sigma$, and a new coupling constant $\gamma^2 =2\pi/k$.
The Euler-Lagrange equations obtained from (\ref{L}-\ref{L_{WZ}}) reproduce
consistently the dynamical equations of the WZNW theory
(\ref{WZNW-Wirkung})
\begin{equation}\label{WZ-Eq}
\partial_{\bar z} (\partial_z g~ g^{-1}) = 0,~~~~~~~~~~~~
\partial_z (g^{-1}\partial_{\bar z} g~) = 0.
\end{equation}
Note that the timelike property of $a$ ($\langle a ~ a\rangle = 1$)
guarantees regularity of (\ref{L_{WZ}}).

\section{The Gauged \slu{} WZNW Theory}

The Lagrangian (\ref{L}-\ref{L_{WZ}}) is invariant under the global
 transformations
\begin{equation}\label{transform}
g \mapsto h_{a}(\varepsilon )g h_{a}(\varepsilon),~~~
\mbox {with}~~~~~~
h_{a}(\varepsilon) = e^{\varepsilon a},
\end{equation}
which for timelike $a$ form the $U(1)$ subgroup of \sll .

By a standard gauging procedure we introduce the
 $U(1)$ gauge
fields $A_z$, $A_{\bar z}$ and get the new Lagrangian
\begin{equation}\label{GI-Lagrang}
{\mathcal L}_G (g,A_z ,A_{\bar z},\partial_z g,\partial_{\bar z}g) =
{\mathcal L}\left(g, \partial_z g -A_z (a g +
g a),\partial_{\bar z}g-A_{\bar z}(ag+ga)\right),
\end{equation}
which is invariant under the local transformations
\begin{equation}
A_z \mapsto A_z +\partial_z \varepsilon ,~~~
A_{\bar z} \mapsto A_{\bar z}+\partial_{\bar z}\varepsilon,~~~
g \mapsto h_{a}(\varepsilon )g h_{a}(\varepsilon ),~~~
\varepsilon = \varepsilon (z,\bar z).
\end{equation}
The non-dynamical gauge fields can easily be eliminated from
(\ref{GI-Lagrang}) through their algebraic equations of motion
\begin{equation}\label{G-fields}
 A_z =\frac{\langle a~\partial_zg~ g^{-1} \rangle}
 {1+\langle a~g~a~g^{-1}\rangle},~~~~~~
 A_{\bar z} =\frac{\langle a~g^{-1}~\partial_{\bar z}g \rangle}
 {1+\langle a~g~a~g^{-1}\rangle}.
\end{equation}
So we obtain a gauge invariant Lagrangian only in terms of the field $g$
\begin{eqnarray}\label{G-Lagrang}
{\mathcal L}_G |&=& -\frac{1}{\gamma^2}~
\Biggl(\langle g^{-1}\partial_z g~ g^{-1}\partial_{\bar z}g
\rangle
 \nonumber \\
&&~~~~~~~-\frac{\langle a~\partial_zg~ g^{-1} \rangle
\langle a~g^{-1}\partial_{\bar z}g\rangle +
\langle a~\partial_{\bar z}g~g^{-1}\rangle
\langle a~g^{-1}~\partial_zg \rangle}
{1+\langle a~g~a~g^{-1}\rangle} \Biggr).
\end{eqnarray}
Without any  loss of generality we assume $a=t_0$.
Since $\langle t_0 g t_0 g^{-1}\rangle =\Lambda_{00} $ is strictly
positive the denominator in (\ref{G-Lagrang}) is never zero.
It is important for the following that this Lagrangian can be
rewritten entirely in terms of the gauge invariant variables
\begin{equation}\label{GIF}
v_1 = \langle t_1~g \rangle ,~~~~~~~~~ v_2 = \langle t_2~ g \rangle.
\end{equation}
The gauge invariance follows from
$e^{\epsilon t_0}\, t_n\, e^{\epsilon t_0}=t_n\,~{\mbox{for}}\,~n=1,2$.
  Introducing the additional variables $v_0 = \langle t_0 g
\rangle$ and $c = -\langle g
\rangle$, we parameterise $g(z, \bar z) \in$ \sll \, as
\begin{equation}\label{q_n}
  g=cI+v^n~t_n= \left( \begin{array}{cr}
  c-v_2&-v_1-v_0\\-v_1+v_0&c+v_2 \end{array}\right),~~~~\mbox{with}~~~
 c^2+v^nv_n=1.
\end{equation}
Inserting this in (\ref{G-Lagrang}), we find that the dependence on
the gauge variant  fields $v_0$ and $c$ is eliminated, and
the reduced Lagrangian becomes
\begin{equation}\label{R-Lagrang}
{\mathcal L}_G |=
 \frac{\partial_zv_1 \partial_{\bar z}v_1 +
\partial_zv_2\partial_{\bar z}v_2}{\gamma^2(1+v_1^2 + v_2^2)}.
\end{equation}
This Lagrangian has a natural complex structure 
in terms of the
 Kruskal coordinates $u=v_1+iv_2$, $\bar u=v_1-iv_2$
of \cite{Witten}. The resulting equation of motion
\begin{equation}\label{eq}
\delz \delzb u = \bar u~
\frac{\partial_zu~\partial_{\bar z}u}{1+u \ub}
\end{equation}
is just  that
obtained in \cite{CMP,MuWe}.

\section{Integration of the Theory by Lagrangian Reduction}

The WZNW equations of motion  (\ref{WZ-Eq}) have the
well-known general solution
\begin{equation}\label{g-chirality}
g(z,\bar z) = g_L(z)~g_R(\bar z),
\end{equation}
where $g_L(z) $, $g_R(\bar z)$
 are arbitrary \sll \, group valued (anti-)chiral functions.

Now let $ g(z,\bar z)$ be a solution (\ref{g-chirality}) which
satisfies the conditions
\begin{equation}\label{r-condition}
\langle t_0~\del_z g_L(z)~ g_L^{-1}(z) \rangle=0~~~~\mbox {and}
~~~~\langle t_0~g_R^{-1}(\bar z)~\del_{\bar z}g_R(\bar z) \rangle =0 .
\end{equation}
Then, due to
(\ref{G-fields}), the set of
functions $g(z,\bar z), A_z(z,\bar z)= A_{\bar z}(z,\bar z)= 0$ form a
solution of the dynamical equations
derived from
 (\ref{GI-Lagrang}). Since the
Lagrangians (\ref{GI-Lagrang}) and (\ref{R-Lagrang}) have in terms of
the gauge invariant fields $v_1$ and $v_2$ the same dynamical
equations (\ref{eq}), the solutions of (\ref{eq}) can be written as
\begin{equation}\label{GIFs}\!\!\!
u(z,\bar z) = \langle (t_1+it_2)~ g_L(z)~g_R(\bar z) \rangle ,
\end{equation}
where the fields $g_L$ and $g_R$ satisfy (\ref{r-condition}).
Equations (\ref{G-fields}) and (\ref{eq}) imply
vanishing field strength
$F_{z\bar z}=\partial_zA_{\bar z}-\partial_{\bar z}A_z=0$.
Then due to gauge invariance, (\ref{GIFs}) describes the general solution of
(\ref{eq}).

We seek these solutions in terms of unconstrained
(anti-) chiral fields. Therefore, we parameterise $g_L$ and $g_R$
as in  (\ref{q_n})
\begin{equation}\label{g_pm}
g_L(z) =c(z) I+v^n(z) t_n,\quad
g_R(\bar z)=\bar c(\bar z) I+{\bar v}^n(\bar z) t_n,
\end{equation}
and introduce for convenience polar coordinates for the chiral fields
\begin{eqnarray}\label{polar-coord}
c &=&R\cos\beta, ~~~~~~~~
v_0=R\sin\beta,\nonumber \\
v_1&=&r\cos\alpha,~~~~~~~~~
v_2= -r\sin\alpha ,
\end{eqnarray}
and similarly for the anti-chiral ones.
The conditions (\ref{r-condition}) lead to $R^2\beta'
-r^2\alpha'=0$, and with $R^2 -r^2=1$ which follows
from (\ref{q_n}), we deduce
 the relations
\begin{equation}\label{R_pm}
R = \sqrt{\frac{\alpha '}{\alpha '-
\beta '}},~~~~~~~~~~
r= \sqrt{\frac{\beta'}{\alpha '-
\beta'}}.
\end{equation}
Here $'$ denotes differentiation.
The insertion of (\ref{g_pm}) and (\ref{polar-coord}) in (\ref{GIFs})
yields finally
 the general solution of (\ref{eq})
\begin{equation}\label{gen-solution}
u(z,\bar z)
=R(z)\bar r(\bar z)e^{i\bar\alpha(\bar z)-i\beta(z)} +
r(z)\bar R(\bar z)e^{-i\alpha(z)+i\bar\beta(\bar z)},
\end{equation}
which is correctly parameterised by the two chiral
and two anti-chiral  functions $\alpha(z) , \,
\beta(z)$,  $\bar \alpha(\bar z) , \,\bar\beta(\bar z)
$.

As a conformal field theory
(\ref{R-Lagrang}) has a traceless energy-momentum tensor $T_{z\bar
  z}=0$, with the chiral component
\begin{equation}\label{T_}
T=T_{z z} = \frac{1}{\gamma^2}~\frac{\partial_z\ub\partial_z u}{1+u \ub} =
 \frac{1}{\gamma^2}~\left (\alpha^{\prime}\beta^{\prime} +
\frac{(\alpha^{\prime\prime}\beta^{\prime}-\beta^{\prime\prime}
\alpha^{\prime})^2}{4\alpha^{\prime}\beta^{\prime}
(\alpha^{\prime}-\beta^{\prime})^2}\right ),
\end{equation}
and similarly for the anti-chiral part $\bar T=T_{\bar z \bar z}$.
A  free-field form of this energy-momentum tensor,
$T(z)={{\phi}'_1}^2(z)+{{\phi}'_2}^2(z)$, can be
obtained by passing to canonical free fields ($k=1,2$)
\begin{equation}\label{psi12-komp}
\psi_k(\sigma,\tau)=\phi_k(z)+\phib_k(\zb),
\end{equation}
where $\phi_k(z)$ and $\phib_k(\zb)$ are chiral and anti-chiral
components, respectively.  The free-field transformation which solves
this problem is given by
\begin{eqnarray}\label{phi_1,2}
e^{- i\alpha}&=&e^{i\gamma\phi_{2}}\frac{2e^{\gamma\phi_{1}}-
e^{-\gamma\phi_{1}} +
2ie^{\gamma\phi_{1}}\Phi}
{\sqrt{\left ( 2e^{\gamma\phi_{1}}-
e^{-\gamma\phi_{1}}\right )^2 +4e^{2\gamma\phi_{1}}\Phi^2}},\\ \nonumber
e^{- i\beta }&=&e^{i\gamma\phi_{2}}\frac{2e^{\gamma\phi_{1}}+
e^{-\gamma\phi_{1}} -
2ie^{\gamma\phi_{1}}\Phi}
{\sqrt{\left (2e^{\gamma\phi_{1}}+
e^{-\gamma\phi_{1}}\right )^2 +4e^{2\gamma\phi_{1}}\Phi^2}}~,
\end{eqnarray}
where $\Phi(z)$ is defined as the integral of
\begin{equation}\label{delPhi}
\delz\Phi(z) =\gamma e^{-2\gamma\phi_{1}(z)}\delz\phi_{2}(z).
\end{equation}
The general solution (\ref{gen-solution}) then takes the form of
a canonical transformation mapping
 \slu{} fields onto free fields
\begin{eqnarray}\label{w-can}
u = e^{i\gamma(\phi_2+\bar \phi_2)}
 \Bigl (e^{\gamma(\phi_1+\bar \phi_1)}(1+\Phi\bar\Phi)
&-&\frac{1}{4}
e^{-\gamma(\phi_1+\bar \phi_1)}\nonumber \\
&+&\frac{i}{2} \bigl(\Phi e^{\gamma(\phi_1-\bar \phi_1)} +
\bar\Phi e^{-\gamma(\phi_1-\bar \phi_1)}\bigr ) \Bigr ).
\end{eqnarray}
In the following  we will impose periodicity in the spatial direction
\begin{equation}
u(\sigma+2\pi,\tau)=u(\sigma,\tau),~~~
\psi_k(\sigma+2\pi,\tau)=\psi_k(\sigma,\tau).
\end{equation}
While the $\psi_k$ are strictly periodic their chiral and anti-chiral
pieces have the following monodromy
\begin{eqnarray}
\phi_k(z+2\pi)=\phi_k(z)+\frac{p_k}{2},
~~~~~\bar \phi_k(\bar z-2\pi)=\bar \phi_k(\bar z)-\frac{p_k}{2},
\end{eqnarray}
where the $p_k$  are momentum zero modes.
With these boundary conditions the integration of (\ref{delPhi}) gives
\begin{equation}\label{Phi}
  \Phi(z)=-\frac{\gamma}{2\sinh(\frac{1}{2}
\gamma p_1)}
  \int^{2\pi}_0  dz'\;\del_{z'}\phi_2(z')e^{
    -\frac{1}{2}\gamma p_1\epsilon_{2\pi}(z-z')- 2\gamma\phi_1(z')}.
\end{equation}
Inserting (\ref{Phi}) into (\ref{w-can})
 exactly reproduces the results obtained in \cite{NuPh99}.

Thus, we have demonstrated that the gauge invariant
Lagrangian reduction indeed yields an
integration method which allows one to solve the equations of motion
of a non-nilpotently gauged WZNW theory in a straightforward manner.

\section{ Integration of the Theory by Hamiltonian Reduction}

The Hamiltonian reduction of the WZNW theory is an alternative method
for the construction and integration of  coset
models. The phase space of the system (\ref{L}) is given by a set of
functions $R(\sigma ), g(\sigma)$, where $R(\sigma )$ and
$g(\sigma )$ take values in the \slua \, algebra and the \sll \, group,
respectively.
The reformulated WZNW action (\ref{L}-\ref{L_{WZ}}) can be written
as $S=\int\left(\theta-H\, \mbox{d}\tau\right)$
where the 1-form $\theta$ and the Hamiltonian $H$ are
\begin{eqnarray}\label{theta}
\theta &=&\int_0^{2\pi} d\sigma\left (
- \langle R g^{-1} \mbox{d} g \rangle
+\frac{\langle t_0 g^{-1}g' \rangle
\langle t_0 \mbox{d} g g^{-1} \rangle  -
\langle t_0 g' g^{-1} \rangle
\langle t_0 g^{-1} \mbox{d} g \rangle }
{\gamma^2(1+\langle t_0~g~ t_0~g^{-1} \rangle ) }\right ) \nonumber
\\
H&=&-\frac{1}{2}\int_0^{2\pi} d\sigma~
\left (\gamma^2~\langle R~R\rangle +
\frac{1}{\gamma^2}~\langle g^{-1}~g'~g^{-1}~g' \rangle \right ).
\end{eqnarray}
Here $~g'=\partial_\sigma g$,
and $\mbox{d}$ denotes the exterior derivative.
Variation of $R(\sigma )$ yields the Hamiltonian equation
\begin{equation}\label{R}
\gamma^2~R(\sigma) = g^{-1} \partial_\tau g.
\end{equation}
Accordingly,
 we parameterise the functions $R(\sigma ),~g(\sigma )$ by the \sll \,
group valued fields $g_L$ and $g_R$
\begin{eqnarray}\label{g_pm(x)}
g(\sigma )&=&g_L(\sigma )g_R(-\sigma),
\nonumber \\
R(\sigma )&=&
g_R^{-1}(-\sigma )g_L^{-1}(\sigma )g_L' (\sigma )g_R(-\sigma )
+g_R^{-1}(-\sigma )g_R'(-\sigma).
\end{eqnarray}
Then the Hamiltonian in (\ref{theta}) splits into chiral and
anti-chiral parts $H =H_L+H_R$, where
\begin{equation}\label{chiral H}~
H_L=-\frac{1}{\gamma^2}\int_0^{2\pi} d\sigma~
\langle g_L^{-1}~g_L'~g_L^{-1}~g_L' \rangle,
\end{equation}
and similarly for $H_R$.
The corresponding splitting holds (up to boundary terms)
also for the symplectic form $\omega =\mbox{d}\theta$. Using ({\ref{theta})
and (\ref{g_pm(x)}) we obtain
\begin{eqnarray}\label{omega}
\omega &=&-\frac{1}{\gamma^2}\int_0^{2\pi} d\sigma~
\left (\langle ( g_L^{-1}~ \mbox{d} g_L)'\wedge
( g_L^{-1}~ \mbox{d} g_L )
\rangle -\langle ( \mbox{d} g_R~  g_R^{-1})'\wedge
 (\mbox{d} g_R~ g_R^{-1}) \rangle \right )
\nonumber \\
&&-\frac{1}{\gamma^2}\left (\langle ( g_L^{-1}~ \mbox{d} g_L )
\wedge
 ( \mbox{d} g_R~  g_R^{-1})
\rangle - \frac{\langle t_0\, g^{-1}\, \mbox{d} g \rangle \wedge
\langle t_0\,\mbox{d}  g \,  g^{-1} \rangle}
{1+\langle t_0~ g~ t_0~ g^{-1} \rangle}\right )
\Bigg{|}_0^{2\pi}.
\end{eqnarray}
The last term of this equation vanishes for periodic $ g$
and in this case (\ref{omega}) reduces to the WZNW symplectic form
of  \cite{CGHOS}.
Then  the Hamiltonian equations split into $\dot{g}_L= g_L',$
and $\dot{ g}_R= -g_R'$, providing the general solution
(\ref{g-chirality}). That is why we  used the same
notation for the $g_{L}$, $g_R$  fields in the Hamiltonian and
 Lagrangian approaches.

The  gauging procedure  which led
to the coset model (\ref{R-Lagrang}) is equivalent to a Hamiltonian
reduction with the same constraints (\ref{r-condition}).  For the
parameterisation (\ref{g_pm}), (\ref{polar-coord}) the reduced
chiral Hamiltonian and 2-form become
\begin{equation}\label{H-theta}
H_L |= \frac{1}{\gamma^2}\int_0^{2\pi} d\sigma~
(f^{{\prime}2} + \alpha'\beta' ), ~~~~
\omega_{L}|=\frac{1}{\gamma^2} \int_0^{2\pi} d\sigma~
(\mbox d f'\wedge \mbox d f  + \mbox d \beta'\wedge \mbox d \alpha ),
\end{equation}
where $\tanh^2f ={\beta'}/{\alpha'}$, and we have neglected boundary
terms in $\omega_L$.
Note that the integrand of $H_L$
is just the energy-momentum tensor (\ref{T_}).

A canonical free-field form of (\ref{H-theta}) can be obtained
by passing to the
canonical free fields $\phi_{1}$ and $\phi_{2}$ using again the
free-field transformation (\ref{phi_1,2}), and we get
\begin{eqnarray}\label{can-form}
T=\frac{1}{\gamma^2}(f'^{2} +\alpha'\beta')&=&\phi_{1}'^{2}+\phi_{2}'^{2},
\nonumber \\
\frac{1}{\gamma^2}(\mbox d f'\wedge\mbox{d}f +\mbox d\beta'
\wedge\mbox{d}\alpha) &=&
\mbox{d}\phi_{1}'\wedge\mbox{d} \phi_{1}+
\mbox{d}\phi_{2}'\wedge\mbox{d}\phi_{2}  \nonumber \\
&&+\mbox {boundary terms}.
\end{eqnarray}
In fact the free-field transformation (\ref{phi_1,2}) was obtained
as a solution of these equations.

This shows that the Hamiltonian reduction like the Lagrangian
reduction provides a convenient approach for the integration of our
non-nilpotently gauged \slu{} WZNW theory.  We envisage that these
methods should be generalisable to  other gauged
WZNW theories.

\section{The Parafermionic \slu{} Coset Currents}

It is well known that the chiral WZNW currents
\begin{equation}\label{Kac}
\gamma^2 J_k(z) = \langle t_k \del_{z}g(z,\zb)~g^{-1}(z,\zb) \rangle =
\langle t_k \del_{z}g_{L}(z)~g_{L}^{-1}(z) \rangle
\end{equation}
satisfy the linear Kac-Moody algebra
\begin{equation}\label{Kac-Moody}
\pk{J_k(z)}{J_l(z')} = {\epsilon_{kl}}^m J_m(z)\delta(z-z')
+\frac{1}{2\gamma^2}
 \eta_{kl} \delta'(z-z').
\end{equation}
 Here we would
like to understand how these properties are impacted by the reduction.
The chiral currents
\begin{equation}\label{V-current}
\gamma^2 V_{\pm}(z) =
 \langle (t_1 \pm it_2)\del_{z}g_{L}(z)~g_{L}^{-1}(z) \rangle
\end{equation}
are of particular interest. Taking into account the parameterisations
(\ref{g_pm},\ref{polar-coord}), the constraints (\ref{R_pm}) and the
free-field transformations (\ref{phi_1,2}), a straightforward
 calculation yields
\begin{equation}\label{V-coset}
V_{\pm}(z) = \frac{1}{\gamma}
\Bigl(\del_z\phi_1(z) \pm i\del_z\phi_2(z)
\Bigr)e^{\pm2i\gamma\phi_2(z)}.
\end{equation}
Interestingly, these fields are the free-field
transformed conserved parafermions of \cite{BCR, MuWe}, which now obtain a
coset current interpretation  through the reduction.

The Sugawara energy-momentum tensor of the WZNW theory
\begin{equation}\label{T-Sugawara}
T(z) =-\gamma^2 J^k(z)J_k(z)
\end{equation}
retains this simple form  in terms of the parafermionic coset
currents \cite{BCR,MuWe} even after   reduction
\begin{equation}\label{T-coset}
 T(z) = \gamma^2 V_{+}(z) V_{-}(z) .
\end{equation}
The free-field parameterisation of the parafermions (\ref{V-coset})
clearly leads to the free-field energy-momentum tensor (\ref{can-form}).
But in contrast to the WZNW Kac-Moody currents (\ref{Kac}), the coset
currents satisfy
a non-linear and non-local Poisson bracket algebra,
 which can be calculated either through the Dirac bracket
method or directly
from the free field representations.
The results for the Poisson brackets depend on the chosen boundary
conditions.
Here we work with  the periodic ones prescribed in section 3.
 For the chiral and
anti-chiral components $\phi_k(z)$, $\phib_k(\zb)$ of the canonical
free fields $\psi_k(\sigma,\tau)$ (\ref{psi12-komp}) we choose the usual mode
expansion
\begin{eqnarray}\label{mode-exp}
\phi_k(z)&=&\frac{1}{2}q_k+\frac{1}{4\pi}p_kz+
\frac{i}{\sqrt{4\pi}}\sum_{n\neq0}
  \frac{a^{(k)}_n}{n}\mbox{e}^{-i nz}, \nonumber\\
 \bar{\phi}_k(\zb)&=&\frac{1}{2}q_k+\frac{1}{4\pi}p_k\zb+
\frac{i}{\sqrt{4\pi}}\sum_{n\neq0}
  \frac{\bar a^{(k)}_n}{n}\mbox{e}^{-i n\bar{z}},
\end{eqnarray}
with the canonical mode algebra
\begin{eqnarray}\label{mode-alg}
\pk{q_k}{p_l}&=&\delta_{k,l}, \,\,\,\,\,\,
i\pk{a^{(k)}_n}{a^{(l)}_n}=m\delta_{k,l}\delta_{m+n,0},
 \nonumber\\
\pk{q_k}{a^{(l)}_m}&=&0,\,\,\,\,\,\,\,~~~~~ \pk{p_k}{a^{(l)}_m}=0.
\end{eqnarray}
Instead
of the $V_\pm(z)$'s we will consider the periodic parafermions
\cite{NuPh99}
\begin{equation}\label{Vpminphi}
  W_\pm(z)=\frac{1}{\gamma}\Bigl(\delz\phi_1(z)\pm i\delz\phi_2(z)
\Bigr)
  e^{\pm2i\gamma\varphi_2(z)},
\end{equation}
where $\varphi_2$ is $\phi_2$ with the momentum zero mode $p_2$
removed and the whole $q_2$ zero mode of the free field
(\ref{psi12-komp}) included, i.e. $\varphi_2(z) = \frac{1}{2}q_2 +
\phi_2(z)|_{p_2=0}$. These periodic coset currents have the
 algebra
\begin{eqnarray}\label{valgebra}
  \{W_\pm(z),W_\pm(z')\}&=&\gamma^2
  W_\pm(z)W_\pm(z')\,h(z-z'),\nonumber\\
  \{W_\pm(z),W_\mp(z')\}&=&-\gamma^2
  W_\pm(z)W_\mp(z')\,h(z-z')\nonumber \\
&&+ \frac{1}{\gamma^2}\left(\partial_z+\frac{i\gamma p_2}{2\pi}\right)
\delta_{2\pi}(z-z'), \nonumber\\
  \{p_2,W_\pm(z')\}&=&\mp2i\gamma W_\pm(z'),
\end{eqnarray}
where
\begin{equation}\label{h}
h(z)=\left(\epsilon_{2\pi}(z)-\frac{z}{\pi}\right)
\end{equation}
is the periodic sawtooth function and $\epsilon_{2\pi}(z)$ the
stairstep function\footnote{$\epsilon_{2\pi}
(\sigma)=2n+1$ for
  $2n\pi<\sigma<(2n+2)\pi$ which coincides with $\hbox{sign}(\sigma)$
  for $-2\pi<\sigma<2\pi$.}.
Note that the momentum zero mode $p_2$ enters into the periodic
parafermion algebra.

The energy-momentum tensor (\ref{T-coset}), now expressed in terms of
the $W_\pm$'s, provides
 the Virasoro algebra
\begin{equation}\label{Virasoro-alg}
 \{T(z),T(z')\}=-\del_{z'}T(z')\delta_{2\pi}(z-z') + 2T(z')\delz\delta_{2\pi}
(z-z'),
\end{equation}
indicating for it conformal weight $\it two$, and the parafermions
$W_\pm(z)$
\begin{eqnarray}\label{primary}
  \{T(z),W_\pm(z')\}&=&-\partial_{z'}  W_\pm(z')\delta_{2\pi}(z-z')+
  W_\pm(z')\delz\delta_{2\pi}(z-z')\\ \nonumber
&&\mp\frac{i\gamma p_2}{2\pi}W_\pm(z')
\delta_{2\pi}(z-z')
\end{eqnarray}
have  conformal weight $\it one$. Finally, we add a useful formula which
generates the energy-momentum tensor $T(z)$ through a Poisson bracket
\begin{eqnarray}\label{empb}
\{D_z W_+(z),W_-(z')\}&=&
\gamma^2 D_z W_+(z)W_-(z')h(z-z') \\ \nonumber
&&
-2T(z)\delta_{2\pi}(z-z')
+\frac{1}{\gamma^2}
D_z^2
\delta_{2\pi}(z-z'),
\end{eqnarray}
where
\begin{equation}
D_z=\partial_z+\frac{i\gamma p_2}{2\pi}.
\end{equation}
Equation (\ref{empb})
 becomes important quantum mechanically because the operator
product $W_{+}(z)W_{-}(z)$ is ill defined and cannot be used to define
a $T(z)$ operator.

\section{ Canonical Quantisation of the Parafermions}

 In this section we shall determine
 explicitly the quantum
analogue of the parafermion algebra (\ref{valgebra})
instead of analysing operator
product expansions. The quantisation of the theory will be defined by
replacing Poisson brackets of the canonical free fields by commutators
$i\hbar\{~,~\} \to [~,~]$, and non-linear expressions in the free
fields will be normal ordered.  But calculations with normal ordered
operators usually yield anomalous contributions.  Such anomalies can
be avoided by quantum mechanically deforming the composite operators of
the theory. Therefore, let us  define the normal ordered parafermion operators
corresponding to (\ref{Vpminphi}) as
\begin{equation}\label{:Vpminphi}
  W_\pm(z)=\frac{1}{\gamma}:\Bigl(\eta\delz\phi_1(z)
\pm i\delz\phi_2(z)\Bigr)
  e^{\pm2i\gamma\varphi_2(z)}:\, ,
\end{equation}
where $\eta$ is a deformation parameter with the classical limit
$\eta=1$.

 First we look for the quantum analogue of the Poisson brackets
(\ref{valgebra}), starting with the simplest example
$\{W_+(z),W_+(z')\}= \gamma^2 W_+(z)W_+(z')h(z-z')$. In the
appendix we have explicitly derived  the relation
\begin{eqnarray}
\label{vpluscomm}
\frac{W_+(z)W_+(z')}{e^{i\hbar\gamma^2 h^+(z-z')}}&-&
\frac{W_+(z')W_+(z)}{e^{-i\hbar\gamma^2h^-(z-z')}}=
 \\ \nonumber
&&\frac{i\hbar}{2\gamma^2}\left(\eta^2-1+\frac{\gamma^2\hbar}{\pi}\right)
:e^{2i\gamma\varphi_2(z)}e^{2i\gamma\varphi_2(z')}:
\delz\delta_{2\pi}(z-z'),
\end{eqnarray}
where
\begin{equation}\label{hpm}
h^\pm(z)=\frac{1}{2}h(z)
\mp \frac{i}{2\pi}\log\left(4\sin^2\frac{z}{2}\right)
\end{equation}
are the positive and negative frequency parts of $h(z)$,
respectively.  The right hand side of (\ref{vpluscomm}) is
proportional to the operator
$:e^{2i\gamma\varphi_2(z)}e^{2i\gamma\varphi_2(z')}:$ which evidently
cannot be rewritten  bilocally
in terms of the parafermions, as is
necessary to have a closed  operator
algebra. However we
can remove the offending term altogether by imposing the restriction
\begin{equation}\label{quantumclosure}
\eta^2-1+\frac{\gamma^2 \hbar}{\pi}=0,~~~~~
\mbox{or}~~~~~\eta=\pm\sqrt{1-\frac{\hbar\gamma^2}{\pi}}.
\end{equation}
The classical limit $\eta =1$ corresponds to the positive square root.
 With this choice we have
\begin{equation}\label{result}
\frac{W_+(z)W_+(z')}{e^{i\hbar \gamma^2 h^+(z-z')}}-
\frac{W_+(z')W_+(z)}{
e^{-i\hbar\gamma^2h^-(z-z')}}=0.
\end{equation}
The quantum relations corresponding to the other
 Poisson brackets of (\ref{valgebra}) are (see
the appendix)
\begin{eqnarray}\label{vminusvminus}
\frac{W_-(z)W_-(z')}{e^{i\hbar\gamma^2h^+(z-z')}}-
\frac{W_-(z')W_-(z)}{e^{-i\hbar\gamma^2h^-(z-z')}}&=&0,\\
\label{:vplusvminus}\frac{W_+(z)W_-(z')}{e^{-i\hbar \gamma^2 h^+(z-z')}}-
\frac{W_-(z')W_+(z)}{
e^{i\hbar\gamma^2h^-(z-z')}}&\!\!\!=\!\!\!&
\frac{i\hbar}{\gamma^2}
\left(
\partial_z+\frac{i\gamma p_2}{2\pi}\right)
\delta_{2\pi}(z-z'),\\
 \label{p2v}
[p_2,W_\pm(z)]&=&\pm2\hbar\gamma W_\pm(z).
\end{eqnarray}
As in the derivation of (\ref{result}) it is necessary to impose
(\ref{quantumclosure}) to eliminate anomalous contributions.

 To check that these operator relations correspond to the classical
Poisson brackets we expand the exact formulae
in powers of $\hbar$, e.g. equation
 (\ref{result}) gives
\begin{equation}
[W_+(z),W_+(z')]-i\hbar\gamma^2
W_+(z)W_+(z')h(z-z')+{ O}(\hbar^2)=0.\end{equation}
Here we have used the splitting relation
 $h(z)=h^+(z)+h^-(z)$, which follows immediately from (\ref{hpm}).

\section{The Energy-Momentum Tensor Operator}

As was mentioned before, to generate the
quantum energy-momentum tensor we should consider the quantum analogue
of (\ref{empb}). From the calculations given at the end of the
appendix it follows that
\begin{eqnarray}\label{emcommutator}
\frac{D_z W_+(z)W_-(z')}
{e^{-i\hbar\gamma^2h^+(z-z')}}&-&
\frac{W_-(z')D_z W_+(z)}{
e^{i\hbar\gamma^2 h^-(z-z')}}=
\frac{i\hbar}{\gamma^2}\left(1+\frac{\hbar\gamma^2}{2\pi}\right){D_z}^2
\delta_{2\pi}(z-z')\\ \nonumber
&&-2i\hbar\eta^2\Biggl (:(\delz\phi_1)^2(z'):+
:(\delz\phi_2)^2(z'):\\  \nonumber
&&~~~~~~~~~~~+\frac{\hbar\gamma}{2\pi\eta}\delz^2\phi_1(z')
+
 \frac{\hbar\gamma^2}{(4\pi\eta)^2} \Biggr ) \delta_{2\pi}(z-z').
\end{eqnarray}
The second entry on the right hand side  just
corresponds to the term
$-2T(z')\delta_{2\pi}(z-z')$ of the classical Poisson
bracket (\ref{empb}) which suggests that
\begin{equation}
T(z)= :(\partial_z\phi_1)^2(z):+
:(\partial_z\phi_2)^2(z):
+\frac{\hbar\gamma}{2\pi\eta}\delz^2\phi_1(z)
+\frac{\hbar\gamma^2}{(4\pi\eta)^2}
\end{equation}
is the free-field energy-momentum tensor of our model with
an additional improvement term.
In the $\sigma$-model
interpretation this improvement term could correspond to a non-perturbative
dilaton \cite{Witten,BS,Tseytlin,DVV}.

Note that the energy-momentum tensor obeys the Virasoro algebra
\begin{eqnarray}\label{qvirasoro}
[T(z),T(z')]&=&-i\hbar \partial_{z'}T(z')\delta_{2\pi}(z-z')
+2i\hbar T(z')\delz\delta_{2\pi}(z-z')\\ \nonumber
&&
-\frac{i\hbar c}{24\pi}(\delz^3+2\del_z)
\delta_{2\pi}(z-z'),
\end{eqnarray}
with central charge
\begin{equation}
c=\hbar\left(2+\frac{3\gamma^2}{\pi\eta^2}\right),
\end{equation}
in agreement with the results of \cite{Witten,centralcharge}.

Classically the parafermions are primary fields of weight one.
Quantum mechanically the commutator
\begin{eqnarray}\label{qprimary}
[T(z),W_+(z')]&=&i
\hbar\left(1+\frac{\hbar\gamma^2}{2\pi}\right)W_+(z')
\delz\delta_{2\pi}(z-z')
\\ \nonumber &&-i\hbar
\partial_{z'}
 W_+(z')\delta_{2\pi}(z-z')+\frac{\hbar\gamma}{2\pi}:p_2W_+(z'):
\delta_{2\pi}(z-z')
\end{eqnarray}
shows that the quantum parafermions have the shifted conformal weight
$1+\hbar\gamma^2/(2\pi)$.

\section{Auxiliary Parafermions}

In order to quantise the canonical transformation
 (\ref{w-can}) we need the operators corresponding to the
simple exponentials in (\ref{w-can})
as well as the non-local fields $\Phi(z)$ and $\bar\Phi(\bar z)$.
However, it turns out that
\begin{equation}\label{APhi}
A(z)=-\frac{1}{2}e^{-2\gamma\phi_{1}(z)} -i\Phi(z),
~~~~~\bar A(\bar z)=-\frac{1}{2}e^{-2\gamma\phi_{1}(\bar z)} -i\Phi(\bar z)
\end{equation}
are more amenable to a quantum treatment. Actually $A(z)$ is one
of the chiral fields
which parametrises the general solution given in
\cite{MuWe}.
The derivative of $A(z)$ has a similar structure to that of the
parafermions, and so we  will refer to $A'(z)$
 as an auxiliary parafermion
\begin{equation}
\tilde V_-(z)=\gamma\Bigl(\partial_z\phi_1(z)
-i\partial_z\phi_2(z)\Bigr)
e^{-2\gamma\phi_1(z)}.
\end{equation}
It  satisfies a closed chiral algebra
if we replace $\phi_1(z)$ by
 $\frac{1}{2}q_1+\phi_1(z)$.
The somewhat artificial doubling of the $q_1$ zero-mode is
 an artifact of our strict
 separation of chiral and anti-chiral objects,
 whereas each term in the general solution
is a product of chiral and anti-chiral pieces.
Following the recipe of section 7 the quantum auxiliary parafermion
turns out to be
\begin{equation}\label{Qauxparaf}
  \tilde V_{-}(z)
 = \gamma:\Bigl(\eta\delz \phi_1(z) - i\delz\phi_2(z)\Bigr)
  e^{-2\gamma\eta^{-1}(\frac{1}{2}q_1+\phi_1(z))}: \, .
\end{equation}
It commutes with $V_-(z')$, but unlike $V_-(z)$ retains its
classical conformal weight {\it one}.  Thus, $A(z)$ has conformal
weight zero, suggesting that it plays the
role of a screening charge.

Summarising, only one fixed deformation
parameter is sufficient for consistent quantisation.
Of course, our quantisation of the \slu{} theory is still incomplete.
In particular, it remains to quantise the black hole metric. It
could answer the question whether a non-perturbative dilaton renders
this metric dynamical.

\vspace{0.5cm}

{\bf Acknowledgments}

\vspace{0.3cm}
G.W. thanks J. Schnittger for discussions on the
quantum aspects  of the problem. We would like to
thank C. J. Biebl for reading the manuscript.  G.J. is grateful to
DESY Zeuthen for hospitality.  His research was supported by
grants from the DFG, GSRT and RFBR (99-01-00151).

\begin{appendix}

\section{Normal Ordered Operator Identities}
\label{distributions}

In this appendix we elaborate on the normal ordered operator
identities quoted in the text. To effect the normal ordering
we decompose the chiral free fields $\phi_i(z)$
defined in (\ref{mode-exp})
as follows
\begin{equation}\label{decomposition}
\phi_i(z)=\frac{1}{2}q_i + \frac{1}{4\pi}p_i z + \phi_i^+(z)+\phi_i^-(z),
\end{equation}
where
\begin{equation}
\phi_i^{\pm}(z)= \pm \frac{i}{\sqrt{4\pi}}\sum_{n>0}
\frac{a_{\pm n}^{(i)}}{n}e^{\mp inz}.
\end{equation}
$\phi_i^-(z)$ and $\phi_i^+(z)$ will be interpreted as creation and
annihilation operators, respectively.
The equivalent anti-chiral constructions
 will not be considered here.

Using the commutator algebra
\begin{equation}\label{mode-com}
[q_{i},p_{j}]=i\hbar\delta_{i,j}, ~~~~~~~~~
[a_m^{(i)},a_n^{(j)}]=m\hbar\delta_{i,j}\delta_{m+n,0},~~~~i,j=1,2
\end{equation}
it follows that
\begin{equation}
[\phi^\pm_i(z),\phi^\pm_j(z')]=0,~~~~~
~~~~~
[\phi_i^\pm(z),\phi_j^\mp(z')]=-\frac{i\hbar}{4}\delta_{ij}h^{\pm}(z-z'),
\end{equation}
where
\begin{equation}
h^\pm(z)=\epsilon^\pm(z)-\frac{z}{2\pi}.
\end{equation}
Here the $\epsilon^\pm(z)$ denote the positive and negative frequency parts
of the stairstep function $\epsilon_{2\pi}(z)$,
and have the Fourier series representation
\begin{equation}\label{epspmsum}
\epsilon^+(z)=
\frac{z}{2\pi}+\frac{i}{\pi}\sum_{n>0}\frac{e^{-in(z-i\varepsilon)}}{n},
~~~~~\epsilon^-(z)=
\frac{z}{2\pi}+\frac{i}{\pi}\sum_{n<0}\frac{e^{-in(z+i\varepsilon)}}{n}.
\end{equation}
Note that we have included a convergence factor, $\varepsilon>0$.  The
$\epsilon^\pm(z)$ functions have the properties
$\epsilon^+{}^*(z)=\epsilon^-(z), ~~ \epsilon^+(-z)=-\epsilon^-(z)$,
and in the limit $\varepsilon\rightarrow 0$ they exhibit
 free-field short
distance singularities
\begin{equation}\label{Ahpm}
\epsilon^{\pm}(z)=\frac{1}{2}
\epsilon_{2\pi}(z)\mp \frac{i}{2\pi}\log\left(4\sin^2\frac{z}{2}\right).
\end{equation}
We also introduce `split' delta functions
$\delta^+(z)=1/2\,\delz\epsilon^+{}(z)$ \cite{for}
\begin{equation}\label{dplus}
\delta^+(z)=
\frac{1}{4\pi}+
\frac{1}{2\pi}\sum_{n>0}e^{-in(z-i\varepsilon)}=
-\frac{1}{4\pi}+\frac{1}{2\pi}\frac{1}{1-e^{-i(z-i\varepsilon)}},
\end{equation}
and similarly  $\delta^-(z)=1/2\,\delz\epsilon^-{}(z)$, which
have the properties
$\delta^+{}^{*}(z)=\delta^-(z)$,
$\delta^+(-z)=\delta^-(z)$, and as
$\varepsilon\rightarrow 0$ the $\delta^\pm(z)$ sum up to the periodic
delta function $\delta_{2\pi}(z)$.

Now we are ready to establish the normal ordered operator
identities quoted in the text. As usual normal ordering
moves creation and annihilation operators respectively to the left and
right, and the Hermitian normal ordering of zero modes $:e^{2q}f(p): =
e^{q}f(p)e^{q}$ will be understood.
With the definitions
\begin{equation}
e_\pm(z):=e^{i\gamma q_2}e^{2i\gamma\phi_2^\pm(z)},~~~
\nu(z):=\frac{1}{\gamma}\Bigl(\eta\delz\phi_1(z)+i\delz\phi_2(z)\Bigr),
\end{equation}
our periodic parafermion operator $W_+(z)$ (\ref{:Vpminphi})
can be written
\begin{equation}
W_+(z)=e_-(z)\nu(z)e_+(z).
\end{equation}
Let us start with a quantum analogue of the Poisson bracket
$\{W_+(z),W_+(z')\}$. Naively one would  consider the commutator,
$[W_+(z),W_+(z')]$, suggesting
that we should compute the operator product
$W_+(z)W_+(z')$.
Using the identity $e^A
e^B= e^B e^A e^{[A,B]}$, which holds if $[A,B]$ commutes with $A$ and
$B$, we have
$e_+(z)e_-(z')=e_-(z')e_+(z)e^{i\hbar\gamma^2h^+(z-z')}$,
so that
\begin{equation}\label{opproduct}
e^{-i\hbar\gamma^2 h^+(z-z')}W_+(z)W_+(z')=
e_-(z)\nu(z)e_-(z')e_+(z)\nu(z')e_+(z').
\end{equation}
But this operator is still not correctly normal ordered.
We decompose $\nu(z)$ as in (\ref{decomposition})
\begin{eqnarray}\label{psipm}
\nu(z)&=&\nu^+(z) +\nu^-(z)
+\frac{\eta p_1+ip_2}{4\pi\gamma},\nonumber \\
\nu^\pm(z)&=&\frac{1}{\gamma}\left(\eta\delz\phi_1^\pm{}(z)+
i\delz\phi_2^\pm{} (z)\right).
\end{eqnarray}
With a little algebra the right hand side of
 (\ref{opproduct}) can be rewritten as follows
\begin{eqnarray}\label{Anti}
e^{-i\hbar\gamma^2h^+(z-z')}\, W_+(z)W_+(z')&=&:W_+(z)W_+(z'):\\ \nonumber
&&+e_-(z)e_-(z')[\nu^+(z),\nu^-(z')]e_+(z)e_+(z')\\ \nonumber
&&+e_-(z)[\nu(z),e_-(z')]\nu(z')e_+(z)e_+(z')\\ \nonumber
&&+e_-(z)e_-(z')\nu(z)[e_+(z),\nu(z')]e_+(z')\\ \nonumber
&&+e_-(z)[\nu(z),e_-(z')][e_+(z),\nu(z')]e_+(z').
\end{eqnarray}
Using the results
\begin{eqnarray}\label{coms}
[\nu(z),e_{\mp}(z')] &=& i\hbar e_{\mp}(z')
\delta^{\pm}(z-z'), \nonumber \\
\,[\nu^+ (z) ,\nu^- (z')] &=&
\frac{i\hbar}{2\gamma^2}(\eta^2 - 1)\delz{\delta^{+}}(z-z'),
\end{eqnarray}
we have
\begin{eqnarray}\label{:opproduct}
\frac{
W_+(z)W_+(z')}{e^{i\hbar\gamma^2 h^+(z-z')}}
&=& :W_+(z)W_+(z'):\\ \nonumber
&&+i\hbar
:\left ( e^{2i\gamma\varphi_2(z)}W_+(z')-
W_+(z)e^{2i\gamma\varphi_2(z')}\right ):\delta^+(z-z')  \\ \nonumber
&&+:e^{2i\gamma\varphi_2(z)}
e^{2i\gamma\varphi_2(z')}:\left(
\frac{i\hbar}{2\gamma^2}(\eta^2-1)\delz\delta^+(z-z')\right. \\ \nonumber
&&\left.~~~~~~~~~~~~~~~~+\hbar^2\left(\delta^+(z-z')\right)^2\right).
\end{eqnarray}
At this point the following
identity is useful
\begin{equation}\label{dplussquared}
\left[\delta^+(z)\right]^2=\frac{1}{(4\pi)^2}+
\frac{i}{2\pi}\partial_z\delta^+(z),
\end{equation}
which is valid even for finite $\varepsilon$.
Using this formula the right hand side
of (\ref{opproduct}) can be written linearly  in $\delta^+(z-z')$ and
its derivative.
Recall that this distribution becomes
 $\delta^-(z-z')$ on exchanging $z$ and $z'$. Thus, if we
take (\ref{:opproduct}) and subtract the equation obtained by
exchanging $z$ and $z'$,
 we get
\begin{eqnarray}\label{preresult}
\frac{W_+(z)W_+(z')}{e^{i\hbar\gamma^2 h^+(z-z')}}&-&
\frac{W_+(z')W_+(z)}{e^{-i\hbar\gamma^2 h^-(z-z')}}=
\\ \nonumber
&&i\hbar
:\left(e^{2i\gamma\varphi_2(z)}W_+(z')-
 W_+(z)e^{2i\gamma\varphi_2(z')}\right):\delta_{2\pi}(z-z')
\\ \nonumber
&&+
\frac{i\hbar}{2\gamma^2}
\left(\eta^2-1+\frac{\hbar
\gamma^2}{\pi}
\right)
:e^{2i\gamma\varphi_2(z)}e^{2i\gamma\varphi_2(z')}:
\delz\delta_{2\pi}(z-z').
\end{eqnarray}
The first term on the right hand side is  zero since the
prefactor of $\delta_{2\pi} (z-z')$ tends to zero as $z\rightarrow
z'$, and so (\ref{vpluscomm}) follows immediately.

Since all the other results quoted in the text can be derived by the
same technique, we will be rather sketchy from now on.
 Evaluating the operator
products $W_+(z)W_-(z')$ and $W_-(z')W_+(z)$
one is led to
\begin{eqnarray}\label{intermediate}
\frac{W_+(z)W_-(z')}{e^{-i\hbar\gamma^2 h^+(z-z')}}&-&
\frac{W_-(z')W_+(z)}{e^{i\hbar\gamma^2 h^-(z-z')}}=\\ \nonumber
&&i\hbar:\Bigl(\nu(z)-\nu^*(z')\Bigr)
e^{2i\gamma\varphi_2(z)}
e^{-2i\gamma\varphi_2(z')}:\delta_{2\pi}(z-z')\\ \nonumber
&&+\frac{i\hbar}{2\gamma^2}\left(\eta^2+1+\frac{\hbar\gamma^2}{\pi}
\right):e^{2i\gamma\varphi_2(z)}e^{-2i\gamma\varphi_2(z')}:
\delz\delta_{2\pi}(z-z').
\end{eqnarray}
Using the identity $f(z)\delta_{2\pi}(z-z')=f(z')\delta_{2\pi}(z-z')$
(valid for periodic $f(z)$)
and derivatives thereof
\begin{eqnarray}\label{:vpm}
\frac{W_+(z)W_-(z')}{e^{-i\hbar\gamma^2 h^+(z-z')}}-
\frac{W_-(z')W_+(z)}{e^{i\hbar\gamma^2 h^-(z-z')}}&=&
-
\frac{\hbar p_2}{2\pi\gamma}
\delta_{2\pi}(z-z')\\ \nonumber
&&+\frac{\hbar}{\gamma}\left(
\eta^2-1+\frac{\hbar\gamma^2}{\pi}\right)
\delz\varphi_2(z')\delta_{2\pi}(z-z')\\ \nonumber
&&+\frac{i\hbar}{2\gamma^2}\left(\eta^2+1+\frac{\hbar\gamma^2}{\pi}
\right)\delz\delta_{2\pi}(z-z').
\end{eqnarray}
(\ref{:vplusvminus}) follows on imposing (\ref{quantumclosure}).

We conclude  by laying the ground for the derivation
 of the
energy-momentum operator. To prove
(\ref{emcommutator}) one first establishes that
\begin{eqnarray}\label{firststep}
\frac{D_z W_+(z)W_-(z')}{e^{-i\hbar
\gamma^2 h^+(z-z')}}-
\frac{W_-(z')D_z W_+(z)}{e^{i\hbar\gamma^2 h^-(z-z')}}
&=&\\
-2i\hbar\gamma^2 A(z,z')
&+&\frac{i\hbar}{\gamma^2}D_z^2
\delta_{2\pi}(z-z'),\nonumber
\end{eqnarray}
where
\begin{equation}
A(z,z')=\frac{ W_+(z)W_-(z')}{e^{-i\hbar
\gamma^2 h^+(z-z')}}\delta^+(z-z')+
\frac{W_-(z') W_+(z)}{e^{i\hbar\gamma^2 h^-(z-z')}}\delta^-(z-z').
\end{equation}
In computing $A(z,z')$ we encounter the same operator products
as in the calculation of (\ref{intermediate}). The result
can be written as
\begin{eqnarray}\label{aobject}
A(z,z')&=&\left(:W_+(z)W_-(z'):+\frac{\hbar}{16\pi^2}
\right)\delta_{2\pi}(z-z')\\ \nonumber
&&-\frac{\hbar}{2\pi}:e^{2i\gamma\varphi_2(z)}
e^{-2i
\gamma\varphi_2(z')}\Bigl(\nu(z)-\nu^*(z')\Bigr):\delz\delta_{2\pi}
(z-z')\\ \nonumber
&&-\frac{\hbar}{8\pi\gamma^2}\left(
\eta^2+1+\frac{\hbar\gamma^2}{\pi}\right)
:e^{2i\gamma\varphi_2(z)}e^{-2i\gamma\varphi_2(z')}:
\delz^2\delta_{2\pi}(z-z').
\end{eqnarray}
Again we used (\ref{dplussquared}) and its derivative
$\delta^+(z)\delz\delta^+{}(z)
=i\delz^2\delta^+{}(z)/(4\pi)$.
With the help of (\ref{quantumclosure}) as well as standard properties
of the delta function
(\ref{aobject})
 reduces to
\begin{eqnarray}\label{laststep}
\!\!\!\!\!\!\! A(z,z')=&&\left[
\frac{\eta^2}{\gamma^2}\left(:(\delz\phi_1)^2(z'):+:(\delz\phi_2)^2(z'):
\right)
+\frac{\hbar\eta}{2\pi\gamma}\delz^2\phi_1(z')\right.
\nonumber\\
&&~~~~~~~~~~~~~~~\left.+\frac{\hbar}{16\pi^2}-\frac{\hbar}{4\pi\gamma^2}
\left(\partial_z+\frac{i\gamma p_2}{2\pi}\right)^2\right]
\delta_{2\pi}(z-z').
\end{eqnarray}
Inserting this into (\ref{firststep}) gives (\ref{emcommutator}).

\end{appendix}


\end{document}